# Ab initio thermodynamic properties of Iridium: A high-pressure and high-temperature study

Balaram Thakur[1*], Xuejun Gong[1,2], and Andrea Dal Corso[1,2]

[1]International School for Advanced Studies (SISSA), Via Bonomea 265, 34136 Trieste, Italy.

[2]CNR-IOM, Via Bonomea 265, 34136 Trieste, Italy.

*Corresponding author: bthakur@sissa.it

**Abstract:**

The high-pressure and high-temperature thermodynamic properties of iridium are studied using density functional theory in combination with the quasi-harmonic approximation, where both the contributions to the free energy of phonons and of electronic excitations are considered. The reliability of different exchange and correlational functionals [Perdew-Burke-Ernzerhof generalized gradient approximation (PBE) (Perdew *et al.* Phys. Rev. Lett. 77, 3865 (1996)), PBE modified for dense solids (PBEsol) (Perdew *et al.* Phys. Rev. B 100, 136406 (2008)) and local density approximation (LDA) (Perdew *et al.* Phys. Rev. B 23, 5048 (1981))], for studying the equation of state (EOS), the phonon dispersions, the mode-Grüneisen parameter, and different thermodynamic properties like thermal pressure, volume thermal expansivity, isobaric heat capacity, bulk modulus, and the average Grüneisen parameter are tested. Elastic constants are studied at $T=0$ K as a function of pressure. The predicted results are compared with the available experiments and previous theoretical data. We find generally a good agreement with experiments with at least one functional, but none of the three outperforms the others in all the investigated thermodynamic properties. The electronic excitations contribution is minimal in bulk modulus, but it is significant for other thermodynamic properties.

Email: Balaram Thakur (bthakur@sissa.it), Xuejun Gong (xgong@sissa.it),
Andrea Dal Corso (dalcorso@sissa.it)

# 1. Introduction:

Only next to the hexagonal close-packed osmium (Os), face-centered cubic (fcc) iridium (Ir) is a second-dense material on the periodic table with the third-highest bulk modulus after diamond and Os [1]. As a refractory material, it is ideal for high-pressure and high-temperature (HPHT) environments. Experimentally, the thermodynamic properties of iridium, like heat capacity, thermal expansion, and bulk modulus, are documented in the Refs. [2–10]. In addition, the shock-wave measurement provides the equation of state (EOS) in the extreme environment, described in the recent database [11] and in Refs. [12–15]. However, only a handful of theoretical reports address the experimental thermophysical properties. These reports, dependent on the methodology and approximations of the calculation, do not always agree among themselves and have varying degrees of accuracy with respect to the experiment.

Recently, Luo *et al*. [16] used first principle molecular dynamics (FPMD) and studied the thermal equation of state using the scalar relativistic Garrity-Bennett-Rabe-Vanderbilt (GBRV) [17] ultrasoft (US) pseudopotential and the Perdew-Burke-Ernzerhof (PBE) [18] functional. Though their EOS agrees satisfactorily with experiments at 300 K, a significant difference is observed at high temperatures. Furthermore, at high temperatures, the authors [16] found a small enhancement in the volumetric thermal expansion coefficient with temperature, which is unusual.

The use of density functional theory (DFT) within the quasi-harmonic approximation (QHA) to accurately study the thermodynamic properties complements the molecular dynamics up to $2/3^{rd}$ of the melting temperature [19]. In this same line, Fang *et al*. [20] used the semi-core Hartwigsen-Goedecker-Hutter (HGH) norm-conserving pseudopotential [21] and the local density approximation (LDA) in their DFT with QHA study of the thermodynamic properties of iridium. However, some of their results, like the temperature-dependent Grüneisen parameter and the variation of thermal pressure ($P_{th}$) with the temperature ($\frac{dP_{th}}{dT}$), which increases with decreasing the volume, contradict Luo *et al*. [16], where $\frac{dP_{th}}{dT}$ decreases with decreasing volume. Fang *et al*. [20] didn't include the electronic excitation contributions (EEC) in the free energy and only considered its role in the isobaric heat capacity. These discrepancies in the literature and the role of electronic excitations on the thermodynamic properties of iridium need to be further investigated.

Moreover, despite having many reports studying some particular property with varying models and approximations, a complete description of the thermodynamic properties of iridium under extreme conditions using the same theoretical method is unavailable. In the present work, we study and compare the anharmonic thermodynamic properties like thermal expansion, isobaric heat capacity, bulk modulus, and thermodynamic average Grüneisen parameter within the QHA using three functionals and consider the role of electronic excitation. Finally, the performance of different functionals under HPHT conditions is discussed.

## 2. Method:

The calculations shown in the present work use `Thermo_pw` [22], a driver of Quantum ESPRESSO [23,24]. The exchange and correlation functional is approximated using the generalized gradient approximation (GGA) proposed by Perdew-Burke-Ernzerhof (PBE) [18], the PBE functional modified for dense solids (PBEsol) [25], and the local-density approximations (LDA) with the Perdew-Zunger (PZ) [26] parameterization. The pseudo-wave functions and the charge densities were expanded in plane waves with kinetic energy cut-offs of 75 Ry and 650 Ry, respectively. Nuclei and core-electrons were treated by the projector augmented wave (PAW) [27] method with pseudopotentials (PPs) from *pslibrary* [28,29]. The iridium atom is described with seventeen valence electrons having a configuration [Xe] $4f^{14}$ $6s^2$ $6p^0$ $5d^7$ and including the $5s^2$ and $5p^6$ semi-core states. For PBE, PBEsol, and LDA, we used the scalar relativistic PPs `Ir.pbe-spn-kjpaw_psl.1.0.0.UPF`, `Ir.pbesol-spn-kjpaw_psl.1.0.0.UPF`, and `Ir.pz-spn-kjpaw_psl.1.0.0.UPF`.

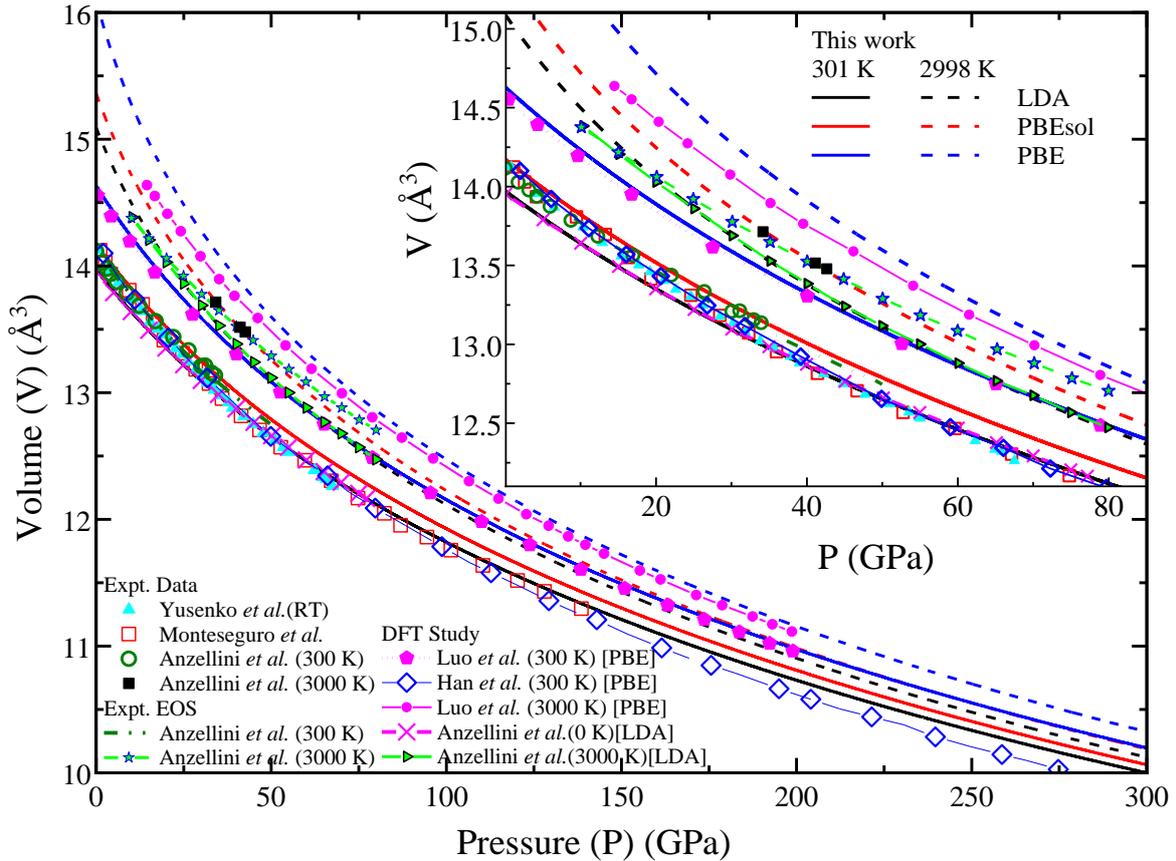

**Fig.1:** P-V equation of state (EOS) of iridium obtained at 301 K and 2998 K using LDA, PBEsol, and PBE functionals. The experimental data from high-pressure compressive measurements by Yusenko *et al.* [30] (at RT), Monteseguro *et al.* [31], and Anzellini *et al.* [32] (at 300 K and 3000 K) are compared. The experimental 3rd-order BM-EOS at 300 K and 3000 K shown by Anzellini *et al.* [32] is displayed. For the DFT study comparison, the EOS of Luo *et al.* [16] (PBE), Han *et al.* [33] (PBE), and Anzellini *et al.* [32] (LDA) are considered. The inset shows a magnified image at low compression.

The iridium's Brillouin zone (BZ) has been sampled on a **k**-point mesh using the Monkhorst-Pack method [34]. The occurrence of the Fermi surface is dealt with by the Methfessel and Paxton (MP) smearing approach [35]. A **k**-point mesh of 32 × 32 × 32 and a MP smearing parameter of σ = 0.02 Ry are used to ensure the calculation's accuracy and efficiency. For phonon calculations, we use density functional perturbation theory (DFPT) [36] extended to PAW [37] on an 8 × 8 × 8 **q**-point mesh. The anharmonic thermodynamic properties are studied under QHA, for which the phonon calculations were performed on a set of 15 volumes with unit cell edge varying from a = 6.1 a.u. to a = 7.5 a.u. (or primitive cell volume from 56.74 a.u.$^3$ (8.408 Å$^3$) to 105.46 a.u.$^3$ (15.628 Å$^3$)) in steps of Δa = 0.1 a.u. The effect of choosing steps of Δa = 0.2 a.u. on the thermodynamic properties of iridium is discussed in supplementary data (S1). The electronic excitations contribution (EEC) is included within the rigid bands' approximation. The procedure to calculate different thermodynamic properties is described in the previous works of our group [38–40].

## 3. Results and discussion:

Fig.1 shows the equation of state calculated for LDA, PBEsol, and PBE at 301 K and 2998 K. The experimental data from high-pressure compressive experiments and the theoretical predictions of different techniques are compared with the present work's EOS. We observe that the DAC experiments carried out at room temperature by Yusenko *et al.* [30] and Montesuguro *et al.* [31] at low pressure (< 20 GPa) agree with the PBEsol EOS. Beyond 20 GPa, the agreement with the LDA EOS is better. Similarly, while comparing our EOS at 301 K and 2998 K, the experimental data and the fitted 3$^{rd}$ order Birch-Murnaghan EOS (3$^{rd}$-BM-EOS) of Anzellini *et al.* [32] determined at 300 K and 3000 K remain close to our PBEsol EOS.

The theoretical prediction made by Luo *et al.* [16] (using PBE) at 300 K and Anzellini *et al.* [32] (using LDA) at 0 K follow well with our EOS computed using PBE and LDA at 301 K, respectively. However, EOS calculated by Han *et al.* [33] at 300 K using a direct integration approach (DIA) to the partition function and PBE functional agree with our PBEsol at 0 GPa and later it decreases with increasing the pressure. To convert Han *et al.* [33] $\frac{V}{V_0}(P)$ results to V(P), we used the experimental V$_0$ of Montesuguro *et al.* [31]*,* for which the predicted results [33] are very close (relative error is - 0.55 %) to the experiment [31] (see Fig. 1). Further, at 3000 K, the predicted EOS by Luo *et al.* [16] tends to follow with our PBE EOS, and interestingly, the EOS computed by Anzellini *et al.* [32] (at 3000 K) and our LDA EOS (at 2998 K) overlap each other.

We fitted the free energy with the 4$^{th}$-order BM equation to calculate the equilibrium lattice constant (a$_0$) from the equilibrium volume (V$_0$), the bulk modulus ($B_T$), and its pressure derivative ($B'_T$). Table 1 compares the obtained parameters with experimental and available theoretical results. At 301 K, relative to the experimental lattice constant [32], our lattice constant is higher by 1.2 % and 0.1 % for PBE and PBEsol, whereas for LDA, it is lower by 0.4 %. Moreover, the present work's lattice constant values at 0 K and 301 K agree with different theoretical studies [32,41,42] (underlined in Table 1, where the values at 300 K are written in italics) having an error below by 0.1% (0.2 %), 0.2 % (0.2 %) and 0.1 % (0.2 %) at 0 K (300 K) for PBE, PBEsol and LDA, respectively.

**Table 1:** Comparison between the equilibrium parameters like lattice constant ($a_o$), bulk moduli ($B_T$), and its pressure derivative ($B'_T$) obtained for different functionals (this work) and other experimental and theoretical studies. The values of the parameters at 0 K and 300 K are shown, where the value at 300 K is in italics. The underlined values are used for the comparison mentioned in the text. The different methods mentioned are PAW-Projected Augmented Wave, ae-all-electron, US-Ultrasoft pp, LAPW-full-potential Linearized Augmented Plane-Wave, and LMTO-Linear Muffin-Tin Orbital.

| Exchange – Correlation Functional | Lattice constant (Å) | | Bulk modulus ($B_T$) (GPa) | | Pressure derivative of $B_T$ ($B'_T$) | |
|---|---|---|---|---|---|---|
| | 0 K | 301 K | 0 K | 301 K | 0 K | 301 K |
| PBE | 3.874 | 3.883 | 348.7 | 336.1 | 5.1 | 5.2 |
| PBEsol | 3.834 | 3.842 | 387.6 | 374.4 | 5.1 | 5.1 |
| LDA | 3.816 | 3.823 | 404.3 | 392.0 | 5.2 | 5.1 |
| Other studies | **Expt (300 K):** 3.837 [31,32] **PBE:** 3.877(PAW) [41], *3.891*(PAW) [42], 3.879(ae) [41], 3.887(LAPW) [43], *3.876*(US) [16]. **PBEsol:** 3.843(PAW) [42], *3.851*(PAW) [42], 3.847(LAPW) [43]. **LDA:** 3.819(PAW) [41], *3.829*(PAW) [32], 3.819(ae) [41], 3.828(FLAPW) [43], 3.815(LMTO) [44]. | | **Expt (300 K):** 360 [32], 339 [31] **PBE:** 342(PAW) [41], 348.8(PAW) [42], 349(ae) [41], *361* (US) [16]. **PBEsol:** 387.9(PAW) [42]. **LDA:** 401(PAW) [41], *377*(PAW) [32], 406(ae) [41], 406.1(PAW) [42], 402.7(LMTO) [44]. | | **Expt (300 K):** 6.0 [32], 5.3 [31] **PBE:** 5.2(PAW) [41], 5.2(ae) [41], *5.3*(US) [16]. **LDA:** 5.2(ae) [41], 5.1(PAW) [41], *5.3*(PAW) [32], 4.8(LMTO) [44]. | |

The equilibrium bulk modulus at 300 K is compared with the experimental studies of Anzellini *et al.* [32] ($P_{max}$ = 35 GPa), and Montesuguro *et al.* [31] ($P_{max}$ = 140 GPa). The bulk modulus comparison (see Table 1) shows that for Anzellini *et al.* [32], the PBE underestimates by 7.1 % (23.9 GPa), whereas PBEsol and LDA overestimate by 3.8 % (14.4 GPa) and 8.2 % (32 GPa), respectively. Instead, for Montesuguro *et al.* [31], the $B_T$ is underestimated by 0.9 % (2.9 GPa) for PBE, and overestimated by 9.5 % (35.4 GPa) and 13.5 % (53 GPa) for PBEsol and LDA, respectively.

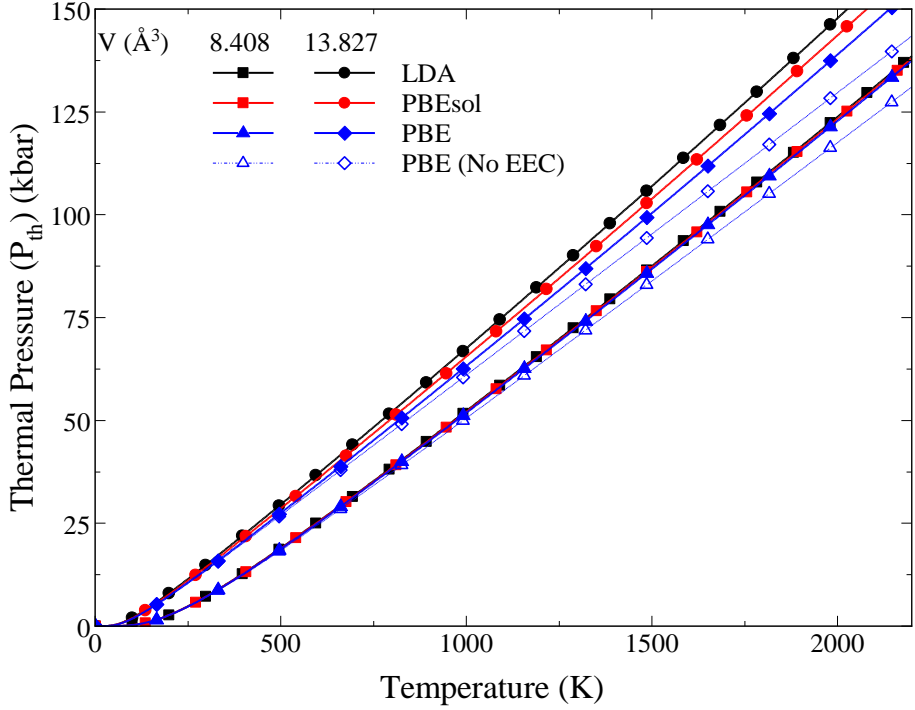

**Fig.2:** Dependence of thermal pressure ($P_{th}$) with the temperature at two volumes 8.408 Å$^3$ and 13.827 Å$^3$ for LDA, PBEsol, and PBE. Filled and empty symbols indicate that the calculation is done while considering and neglecting the electronic excitation contributions (EEC). The variation of $P_{th}$ with no electron excitation contribution (no EEC) is shown only for PBE.

The variation of thermal pressure ($P_{th}$) with temperature is displayed in Fig.2. Here, we studied the $P_{th}$ on the two same-volume used in the present QHA calculations, 8.408 Å$^3$ (the smallest volume) and 13.827 Å$^3$ (nearest to the equilibrium volume). Unlike Fang *et al.* [20], we observe that at a given temperature, the thermal pressure increases with increasing volume, consistent with Luo *et al.* [16]. Comparing the three functionals, we notice that for the system near the equilibrium, PBEsol lies in between PBE and LDA, whereas for the compressed system, all three functionals collapse on a single curve. Furthermore, the EEC on the thermal pressure is studied for each functional, and the dependence of $P_{th}$ with temperature without EEC is shown for PBE in Fig.2. At 2000 K, we found that without EEC, the $P_{th}$ decreases uniformly by ~ 4 % for V = 8.408 Å$^3$ and ~ 7 % for V = 13.827 Å$^3$ for all three functionals.

The variation of volume thermal expansion coefficient (β) with the temperature at 0 GPa and 300 GPa is presented in Fig.3. With respect to LDA, the differences of PBEsol and PBE increase with temperature. The EEC to the β at 0 GPa (see Fig.3) are similar for PBEsol and LDA, whereas it is higher for PBE. At 301 K, the difference between LDA and PBEsol (PBE) for β (in 10$^{-6}$ K$^{-1}$) is 0.62 (2.77), and at 1999 K, it is 1.29 (6.87). Instead, at 301 K and at 0 GPa, the difference of the EEC between LDA and PBEsol (PBE) (in 10$^{-6}$ K$^{-1}$) is 0.02 (0.08), whereas at 1999 K, it is 0.32 (1.54). Also, for PBEsol, at 301 K, with increasing the pressure from 0 GPa to 300 GPa, the EEC decreases by ~ 85 % (from 0.3671 to 0.05511 in 10$^{-6}$ K$^{-1}$), whereas at 1999 K, EEC is reduced by ~ 90 % (from 4.57921 to 0.43535 in 10$^{-6}$ K$^{-1}$). Thus, the EEC effect alone does not explain the difference among the values of β obtained for different functionals. These differences vanish with increasing pressure where the β for different functionals overlap, as shown in Fig.3.

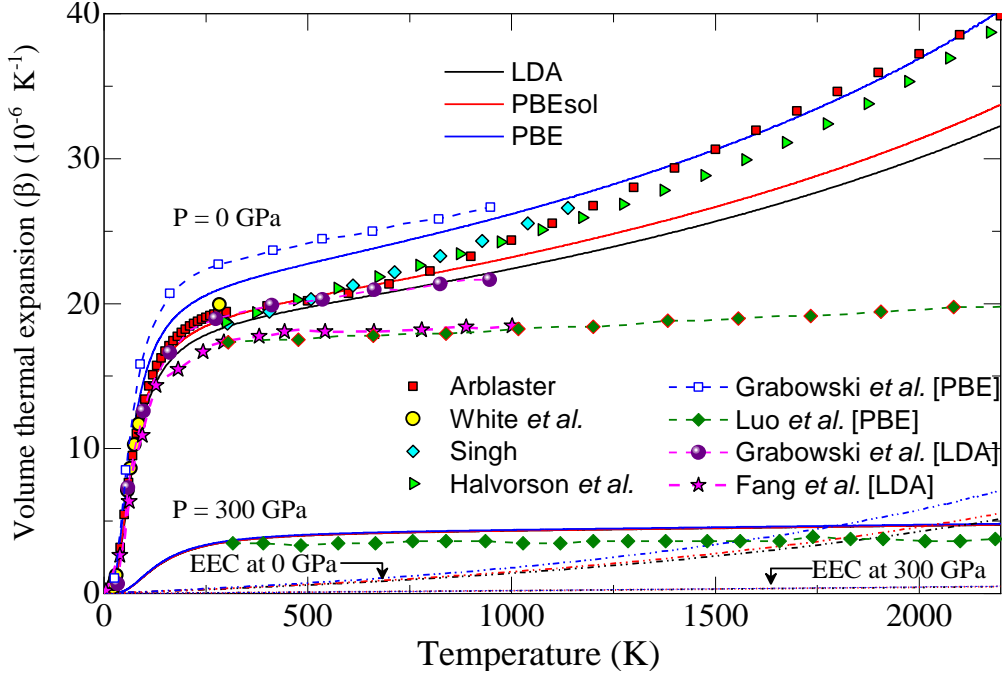

**Fig.3:** Temperature-dependent volumetric thermal expansion (β) obtained for LDA, PBEsol, and PBE at 0 GPa and 300 GPa. The electronic excitation contribution (EEC) to the β at 0 GPa, and 300 GPa is shown. The experimental data (symbols) from Arblaster [6], White *et al.* [7], Singh [8], and Halvorson [9] are included. The simulated results (dash and symbols) from Grabowski *et al.* [41] (PBE and LDA), Luo *et al.* [16] (PBE), and Fang *et al.* [20] (LDA) are considered for comparison.

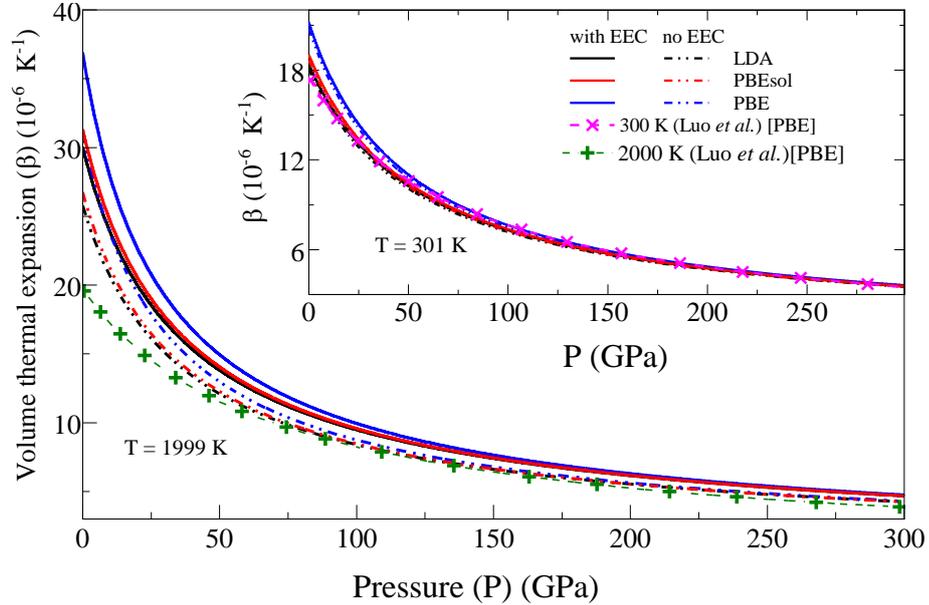

**Fig.4:** Pressure-dependent β at 1999 K and 301 K (inset) calculated with including (solid line) and excluding (dash-dot-dot) EEC for three functionals is displayed. The simulated results (dash and symbols) of Luo *et al.* [16] (PBE) at 2000 K and 300 K are compared.

On comparing with the experimental data from Arblaster [6], White *et al.* [7], Singh [8], and Halvorson *et al.* [9], it is observed that at low temperatures (< 1000 K), the experimental data follows well with PBEsol. In contrast, the PBE functional seems to be a better choice at higher temperatures.

The LDA result of Grabowski et al. [41] is in good agreement with our work, while their [41] PBE is overestimated. Fang et al. [20] (LDA) and Luo et al. [16] (PBE) agree with our curves only at low temperatures (< 300 K), whereas, with increasing the temperature, their predicted values are significantly lower than the experiment and of our results. However, at 300 GPa, our result agrees with Luo et al. [16]. It is noted that Fang et al. [20] and Luo et al. [16] neglected the EEC in the free energy and hence in β. In part for this reason, at 0 GPa, Fang et al. [20] and Luo et al. [16] disagree with our results and other experimental data.

Fig.4 presents the pressure-dependent β obtained for the different functionals at 301 K and 1999 K, considering and neglecting the EEC. At 301 K (inset of Fig.4), the difference between the β obtained with and without EEC is minimum and becomes negligible with increasing pressure, whereas at 1999 K, the difference in β first decreases but does not vanish at higher pressure. The pressure-dependent β calculated in the present work using PBE agrees well with the result of Luo et al. [16] (PBE) at 300 K, even neglecting the EEC. In contrast, at 2000 K, there is a significant difference between ours and Luo et al. [16] results at low pressures, which decreases when the pressure is increased

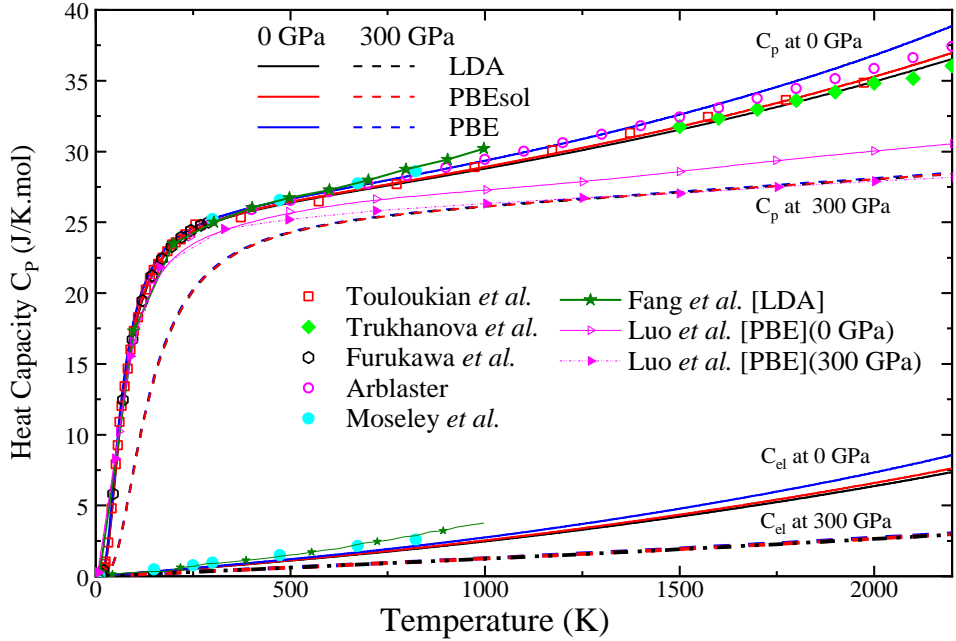

**Fig.5:** (a) Temperature-dependent isobaric heat capacity ($C_P$) and the EEC to the heat capacity ($C_{el}$) at 0 GPa and 300 GPa are presented. The experimental data (symbols) compared are from Touloukian et al. [2] TPRC data series, Trukhanova et al. [3], Furukawa et al. [4], Arblaster [5], and Moseley et al. [45]. Simulated results from Luo et al. [16] (PBE) and Fang et al. [20] (LDA) are included.

Fig.5 shows the temperature-dependent isobaric heat capacity ($C_P$) and the electronic excitation contributions (EEC) to the heat capacity $C_{el}$ for different functionals at 0 GPa and 300 GPa. We noticed that at 0 GPa, the variation of $C_P$ with T for three functionals is identical up to 1000 K and beyond this, the difference between the PBE and other functionals increases with temperature. A similar trend of $C_{el}$ at 0 GPa for different functionals is also observed. At 2000 K relative to LDA, the $C_P$ (in J/K.mol) of PBEsol (PBE) is 0.35 (1.87) larger, whereas the $C_{el}$ (in J/K.mol) is 0.22 (0.97) more. Therefore, the difference in $C_P$ cannot be ascertained totally to the EEC and is in part due to the presence of higher β for PBE with respect to the PBEsol and LDA, as seen in Fig.3.

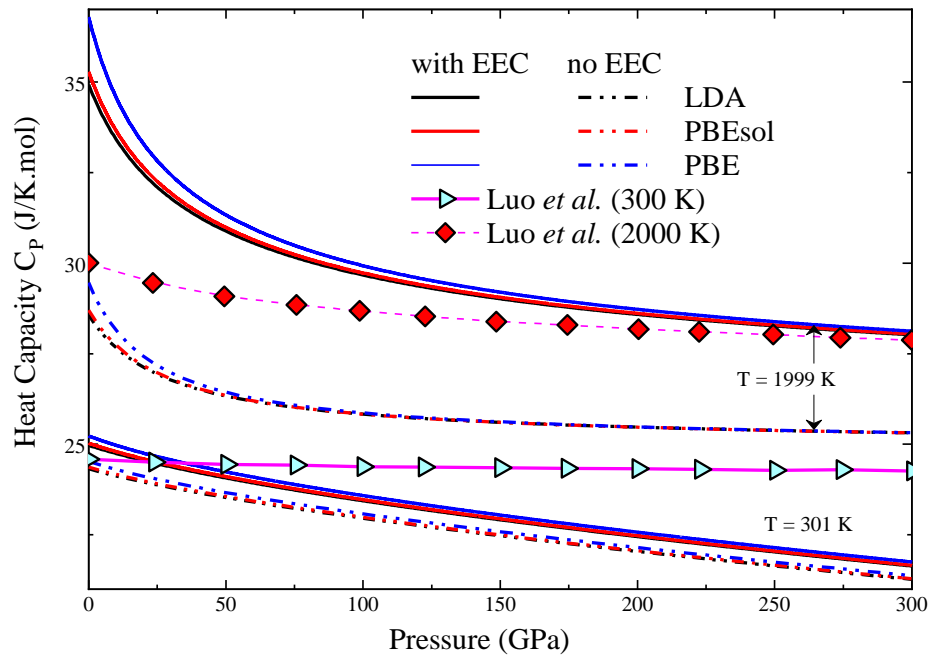

**Fig.6:** Pressure-dependent $C_P$ at 1999 K and 301 K (inset) calculated with including (solid line) and excluding (dash-dot-dot) EEC for three functionals is shown. The simulated result from Luo et al. [16] (PBE) is compared.

Compared with older literature, Touloukian et al. [2] TPRC data series, Trukhanova et al. [3], Furukawa et al. [4], and Arblaster [5], it is observed that our theoretical prediction of $C_P$ agrees well with PBEsol and LDA functionals in the entire temperature range. The $C_P$ and EEC ($C_{el}$) calculated using LDA by Moseley et al. [45] and Fang et al. [20] agree with the present work (see Fig.5). Luo et al. [16] (PBE) at 0 GPa agrees with our simulated results and with experiment only at low temperatures (< 300 K), whereas at 300 GPa, Luo et al. [16] are consistent with our result only for high temperatures. Moreover, in Fig.5, it is seen that at 300 GPa, the $C_P$ and $C_{el}$ for the different functional overlaps and their values (for PBE in J/K.mol) at 2000 K decrease significantly from their 0 GPa values by 14.67 and 4.61, respectively.

The rate of decrease in $C_P$ ($dC_P/dP$) with an increase in pressure depends on the temperature, as illustrated in Fig.6, where from 0 GPa to 300 GPa (say for PBE), the $dC_P/dP$ (in J/K.mol. GPa) is about 0.01 and 0.03 at 301 K and 1999 K, respectively. The difference in the value of $C_P$ at 301 K with and without considering the EEC remains the same. In contrast, at 1999 K, the $C_P$ without EEC decreases initially and remains constant beyond 150 GPa. Similar to Fig.5, at low temperature, Luo et al. [16] (PBE) is consistent with our result at low pressure, while at high temperature, it is consistent only at high pressure (see Fig. 6).

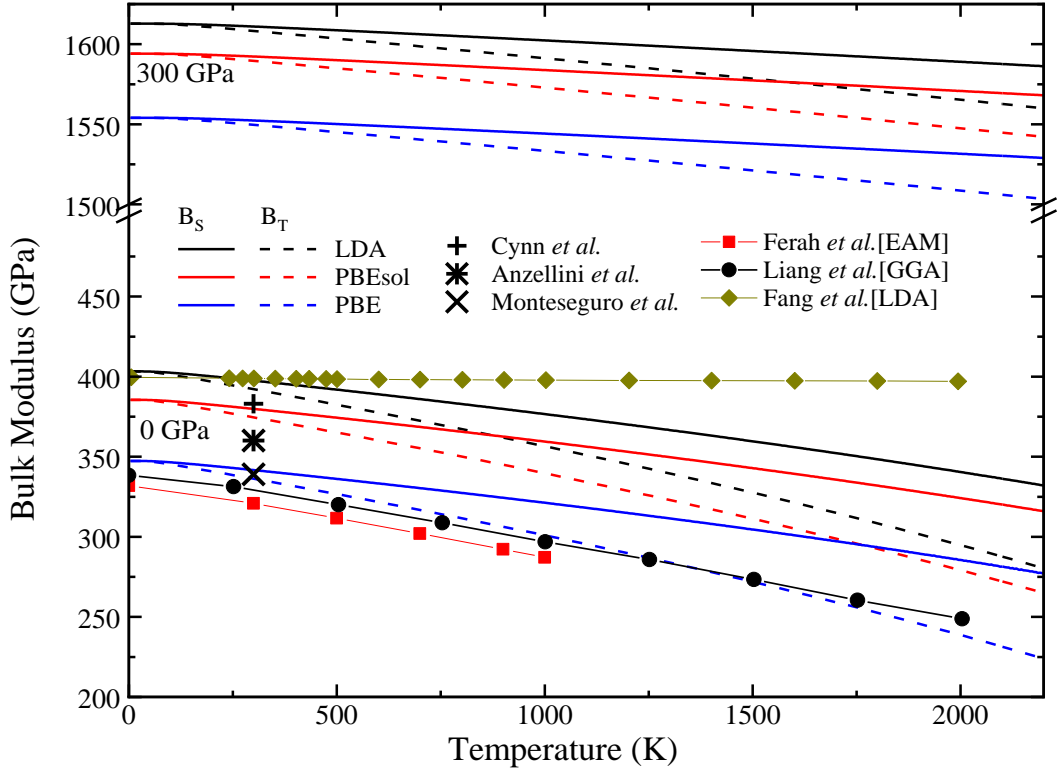

**Fig.7:** Variation of isoentropic ($B_S$) and isothermal ($B_T$) bulk modulus with temperature at 0 and 300 GPa for LDA, PBEsol, and PBE (shown in solid and dashed lines). Comparison with the experimental data at 300 K of Cynn *et al.* [44], Anzellini *et al.* [32], and Montesuguro *et al.* [31] are shown as symbols and theoretical predictions by Ferah *et al.* [46] (electron embedded method (EAM)-molecular dynamics calculations), Liang *et al.* [47] (GGA) and Fang *et al.* [20] (LDA) are represented as a line with symbols.

The isoentropic ($B_S$) and the isothermal ($B_T$) bulk moduli for the three functionals at 0 and 300 GPa are reported in Fig.7, which shows that, with temperature the bulk moduli $B_i$ ($B_S$ and $B_T$) decreases. This decrease in $B_i$ depends on the pressure. For all the functionals, at 0 GPa, the decrease in $B_S$ and $B_T$ ($dB_i/dT$) from 0 K to 2000 K is ~ 0.3 GPa/K and ~ 0.5 GPa/K, respectively, whereas at 300 GPa, the $dB_i/dT$ becomes ~ 0.1 GPa/K and ~ 0.2 GPa/K, respectively. It is noted that on neglecting the EEC (not shown in Fig. 7), an insignificant change in the $B_S$ and $B_T$ is observed.

At 300 K, the experimental bulk modulus studied by Cynn *et al.* [44] is in agreement with the PBEsol, the one of Montesuguro *et al.* [31] is with the PBE, and the one of Anzellini *et al.* [32] lies in the mid-way of PBEsol and PBE. Our predictions and others experimental results overestimate the bulk modulus obtained from the MD simulation by Ferah *et al.* [46]. Moreover, the B(T) calculated by Liang *et al.* [47] (GGA) is in good agreement with our predicted PBE $B_T$. Also, the theoretical value of $B_0$ by Fang *et al.* [20] (LDA) matches well with our LDA curves at up to 500 K; however, their values remain constant with temperature instead of decreasing.

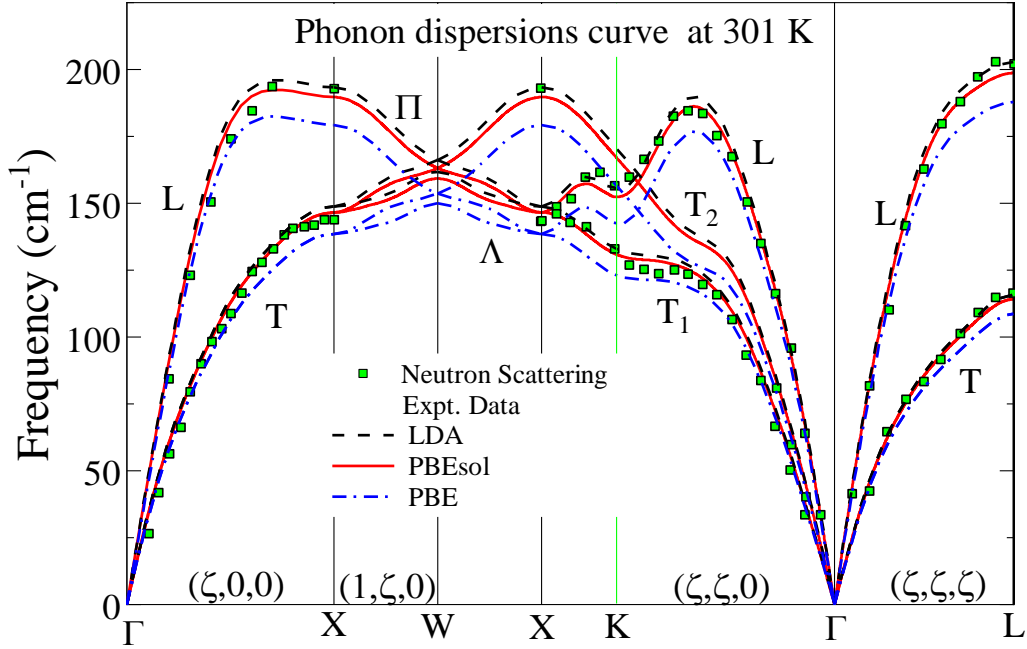

**Fig.8:** Phonon dispersions curve obtained at 301 K for LDA, PBEsol, and PBE (lines) is compared with the room temperature inelastic neutron scattering [48] data (symbols).

In Fig.8, we compare the phonon dispersions measured using the inelastic neutron scattering [48] and calculated using PBE, PBEsol, and LDA. A comparison between phonon dispersion obtained using different **k**-point and **q**-point mesh for PBEsol is discussed in supplementary data (S2). The phonon frequencies shown here are interpolated at 301 K, starting from those calculated on the 15 geometries. The dispersion agrees with Ref. [42]. Also, from Fig.8, we note that the phonon frequencies obtained using the PBEsol and LDA are satisfactory, but for PBE, significant softening is observed. The error in the phonon frequency is well correlated with the error in the lattice parameter rather than the bulk modulus, as discussed earlier [42].

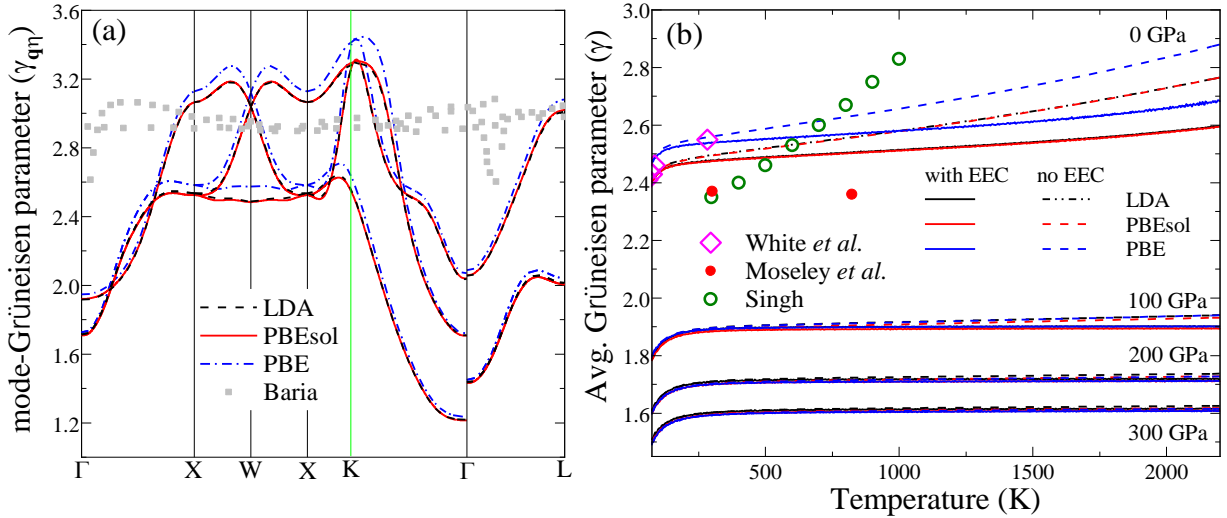

**Fig.9:** (a) Variation of the mode-Grüneisen parameter ($\gamma_{q\eta}$) for LDA, PBEsol and PBE functionals. The theoretically predicted data of Baria [49] are shown, and (b) the variation of a thermodynamic average-Grüneisen parameter as a function of temperature with considering (solid line) and ignoring (dashed) the electron excitation contribution (EEC) at 0 GPa, 100 GPa, 200 GPa, and 300 GPa. The compared values (symbols) are from White *et al.* [7], Moseley *et al.* [45], and Singh [8].

In Fig.9(a), we compare the mode-Grüneisen parameter ($\gamma_{\mathbf{q}\eta}$) obtained for LDA, PBEsol, and PBE. The $\gamma_{\mathbf{q}\eta}$ shows how the phonon modes are affected by the anharmonicity of the system. The non-analyticity at $\Gamma$ is an inherent property of $\gamma_{\mathbf{q}\eta}$, which arises due to the direction dependence of the phonon modes. In Fig.9(a), $\gamma_{\mathbf{q}\eta}$ are similar for LDA and PBEsol, while they are higher for PBE. For comparison, the $\gamma_{\mathbf{q}\eta}$ calculated by Baria [49] are shown. A significant difference between the present study and Baria [49] is observed. Baria [49] studied the iridium's static and vibrational properties using a model pseudopotential, which incorporates the *s-d* hybridization effects and depends only on the effective core radius.

Fig.9(b) compares the temperature variation of the thermodynamic average Grüneisen parameter ($\gamma = \frac{\beta B_T V}{C_V}$) at 0 GPa, 100 GPa, 200 GPa and 300 GPa for the three functionals. The pressure derivatives of $\gamma$ decrease with increasing pressure. Additionally, $\gamma$ remains constant with temperature when the EEC is considered. In contrast, ignoring the EECs leads to the temperature dependence on $\gamma$, especially at low pressures. Also, from Fig.9(b), it is noticed that at 0 GPa, the $\gamma$ from PBEsol and LDA are similar, whereas, for PBE, the value is higher. The separation between the values of $\gamma$ vanishes with pressure, where all three functionals collapse on each other. A similar trend was discussed previously for the thermal expansion coefficient.

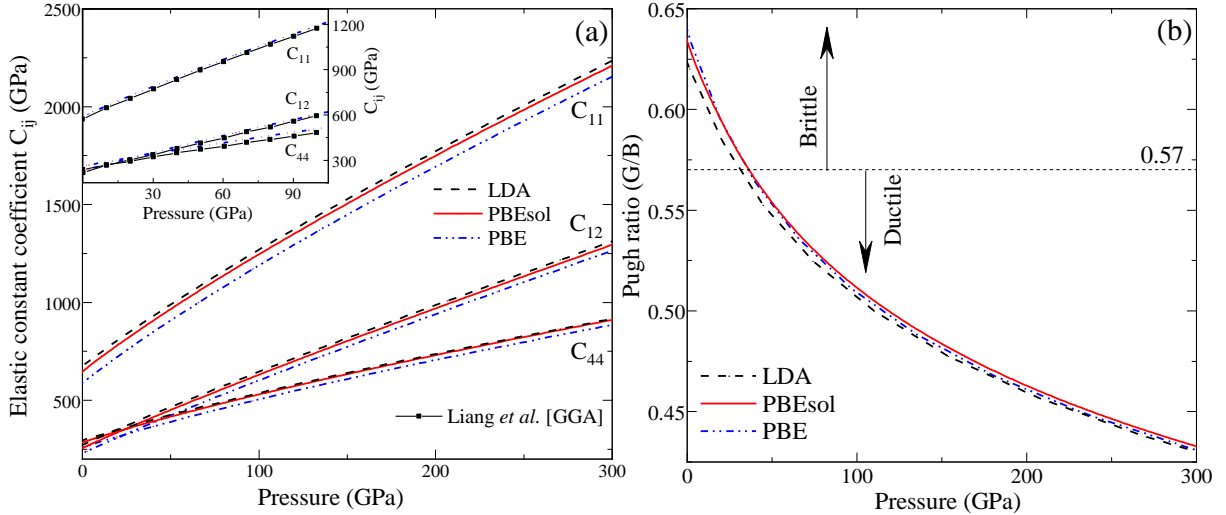

**Fig.10:** (a) Variation of elastic constant coefficient ($C_{ij}$) $C_{11}$, $C_{12}$, and $C_{44}$ with pressure at 0 K, obtained using LDA, PBEsol, and PBE. Inset shows the comparison between $C_{ij}$ obtained using the PBE and with Liang *et al.* [47] (GGA) predicted data (b) Variation of Pugh ratio (G/B), a ratio of shear moduli (G) and bulk modulus (B), with pressure at 0 K for LDA, PBEsol, and PBE. The values of G and B are obtained using Voigt-Reuss-Hill approximation. The horizontal dotted line is at G/B = 0.57, above (or below); the system is considered brittle (or ductile).

On comparing with the experimentally obtained $\gamma$, the result of White *et al.* [7] is found to follow well with the PBE results of the present work. The $\gamma$ value calculated by Moseley *et al.* [45], using DFT with ultrasoft pseudopotentials and LDA, remains reasonably constant. When compared, we found that our values (for LDA) are overestimated by 4.2 % at 300 K and 6 % at 823 K. The overestimation is expected to be due to different technical ingredients. Further, the result of Singh [8] deviates significantly from our results. Singh [8] calculated the value of γ with a fixed value of the volume V and of the compressibility at room temperature, while the thermal expansion coefficient (β) and isochoric heat capacity ($C_V$) vary with temperature. For this, we believe that the γ value of Singh [8] increases linearly with temperature.

**Table 2:** Comparison of equilibrium elastic constant coefficient $C_{ij}$ (in GPa) at 0 K.

|  |  | $C_{11}$ | $C_{12}$ | $C_{44}$ |
|---|---|---|---|---|
| This Study | LDA | 673.8 | 270.9 | 294.6 |
|  | PBEsol | 648.1 | 256.8 | 286.0 |
|  | PBE | 586.7 | 228.6 | 258.3 |
| Expt. Study [50] |  | 596.0 | 252.0 | 270.0 |

In Fig.10, we report the pressure dependence on the elastic constants. Three independent elastic coefficients, $C_{11}$, $C_{12}$, and $C_{44}$ (collectively assigned as $C_{ij}$), are present for fcc iridium with cubic symmetry. For each functional, the $C_{ij}$'s are computed for the 15 different geometries used in QHA. The variation of $C_{ij}$ with the pressure is shown in Fig.10, confirming that for the entire range of geometries studied in the present work, the system satisfies the Born stability criteria [51,52] ($C_{11} > 0$, $C_{44} > 0$, $C_{11} - C_{12} > 0$, and $C_{11} + 2C_{12} > 0$) and therefore all the studied geometries are mechanically stable. The inset of Fig.10 shows that the $C_{ij}$'s calculated using the PBE functional is in good agreement with the DFT study by Liang *et al.* [47] using the GGA. Further, comparing with the experiment [50] at 0 K, the equilibrium $C_{ij}$ obtained using PBE shows better results than PBEsol while LDA is worse among the three functionals (see Table 2).

Moreover, the nature of the material, whether brittle or ductile, can be studied by studying the Cauchy pressure [53], defined as ($C_{12}$–$C_{44}$), and the Pugh ratio (G/B) [54], where G and B are the shear and bulk modulus, respectively. For cubic crystal structures, these two criteria are identical [55]. The material behaves ductile (brittle) if the Pugh ratio is less (more) than 0.57. The Pugh ratio at ~ 0 GPa for LDA, PBEsol, and PBE are 0.62, 0.63, and 0.64, which agrees with the experimental value of 0.63 at 0 K [50] and with theoretical prediction by Liang *et al.* [47] (GGA). The variation of the Pugh ratio with pressures shown in Fig. 10(b) indicates that iridium is brittle at equilibrium conditions, and at 0 K, the transition from brittleness to ductility occurs at an average pressure of ~ 343 kbar (317 kbar, 357 kbar and 355 kbar for LDA, PBEsol, and PBE, respectively). The transition pressure for PBEsol and PBE is considered the same within the tolerance limit.

## 4. Conclusion:

We studied the thermodynamic properties of iridium using the DFT within the QHA, where both the phonon and electronic excitation contributions in the free energy are considered. Three popular functionals, LDA, PBEsol, and PBE, were tested on the thermodynamic properties. Comparison of different properties with experiment and other theoretical models shows good agreement. The lattice constant obtained using PBEsol gives the minimum error of 0.1 %, whereas PBE overestimates the experimental value by 1.2 %, and LDA underestimates of 0.4 %. PBEsol and LDA functionals agree well with the experimental results for phonon dispersions, isobaric heat capacity and low-temperature thermal expansion calculation, whereas the PBE functional is suitable for high-temperature thermal expansion and bulk modulus.

The electronic excitation's role is minimal for bulk modulus, and the contributions are independent of the functional choice for thermal pressure calculations. This contribution in the thermodynamic properties like thermal expansion coefficient, heat capacity, and the thermodynamic average Grüneisen parameter is crucial, particularly at low-pressure and high-temperature conditions. The electronic contributions from PBEsol and LDA are similar, whereas PBE is slightly higher. However, this difference diminishes significantly under high pressure. The elastic constants study at 0 K indicates iridium is brittle at equilibrium conditions, with a transition pressure of ~ 343 kbar from brittle to ductile, in agreement with Liang *et al.* [47]. Finally, we have shown the mode-Grüneisen parameters mapped on the Brillouin zone and the thermodynamic average Grüneisen parameter as a function of temperature and pressure, which are crucial to understand the anharmonicity in iridium.

It is to be noted that the functionals chosen here are simple and efficient, but our results could be further improved if a single functional could explain accurately all the thermodynamic properties. There are several possibilities to choose, from a self-interaction corrected (SIC) functional or hybrid functionals or functionals within the DFT such as a strongly constrained and appropriately normed (SCAN) functional, or also a Hubbard corrected (DFT+U) functional. Using these more sophisticated schemes would be very interesting; however, these corrections require heavy calculations and are computationally costly. For many of them, phonon calculations have not yet been implemented. They must, therefore, be reserved for future research.

**Acknowledgement:**

Computational facilities were provided by SISSA through its Linux Cluster and ITCS and SISSA-CINECA 2021-2024 agreement. This work has been supported by the Italian MUR through the National Centre for HPC, Big Data, and Quantum Computing (grant No. CN00000013).

**Data availability statement:**

All data that support the findings of the present study are included within the article.

Supplementary Data

**Ab initio thermodynamic properties of Iridium: A high-pressure and high-temperature study**

Balaram Thakur[1*], Xuejun Gong[1,2], and Andrea Dal Corso[1,2]

[1]International School for Advanced Studies (SISSA), Via Bonomea 265,34136 Trieste, Italy.

[2]CNR-IOM, Via Bonomea 265, 34136 Trieste, Italy.

*Corresponding author: bthakur@sissa.it

Email:   Balaram Thakur (bthakur@sissa.it), Xuejun Gong (xgong@sissa.it),
         Andrea Dal Corso (dalcorso@sissa.it)

**S1: Comparison of thermodynamic properties of iridium obtained using Δa = 0.1 a.u. (15 geometries) and Δa = 0.2 a.u. (7 geometries) in quasi-harmonic approximation:**

The main article evaluates the thermodynamic properties like the thermal expansion coefficient by differentiating the free energy (F) considered on 15 geometries (i.e. Δa = 0.1 a.u.). Here, we assessed the free energy starting from the second geometry and doubling the step (Δa = 0.2 a.u.), resulting in 7 geometries. The comparison between the free energy (F), volume thermal expansion (β), isobaric heat capacity ($C_P$), bulk modulus (B), and thermodynamic average Grüneisen parameter (γ) obtained from 15 and 7 geometries for PBEsol are shown in Fig.S1(A-E).

- Free Energy:

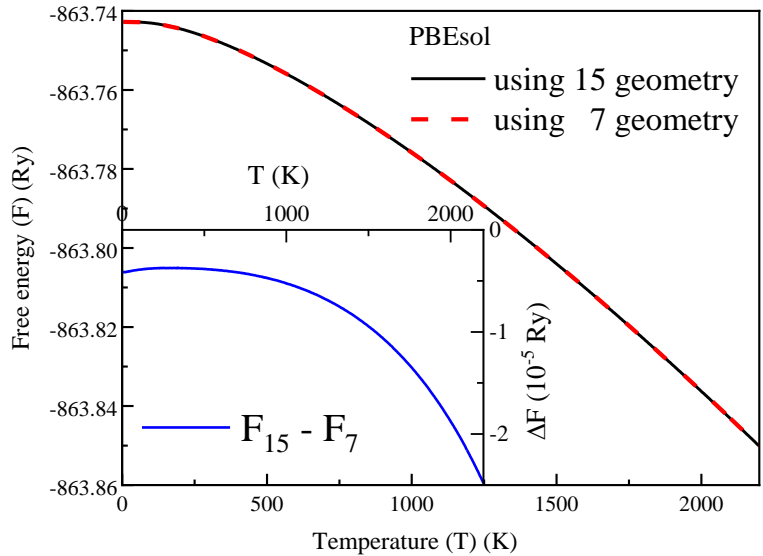

**Fig.S1A:** Temperature-dependent free energy (F) for PBEsol obtained using 15 and 7 geometries. The inset shows the difference ($F_{15} - F_7$) in the free energy for the two cases.

- Volume thermal expansion (β):

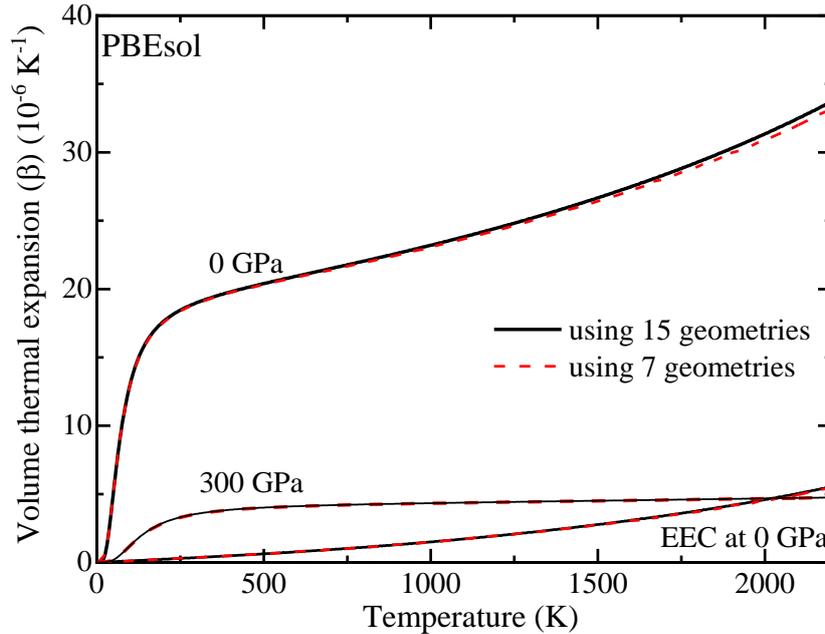

**Fig.S1B:** Temperature-dependent volumetric thermal expansion (β) obtained for PBEsol at 0 GPa and 300 GPa using the derivative of free energies evaluated on 15 and 7 geometries. The electronic excitation contribution (EEC) to the β at 0 GPa is shown.

- Isobaric heat capacity ($C_P$):

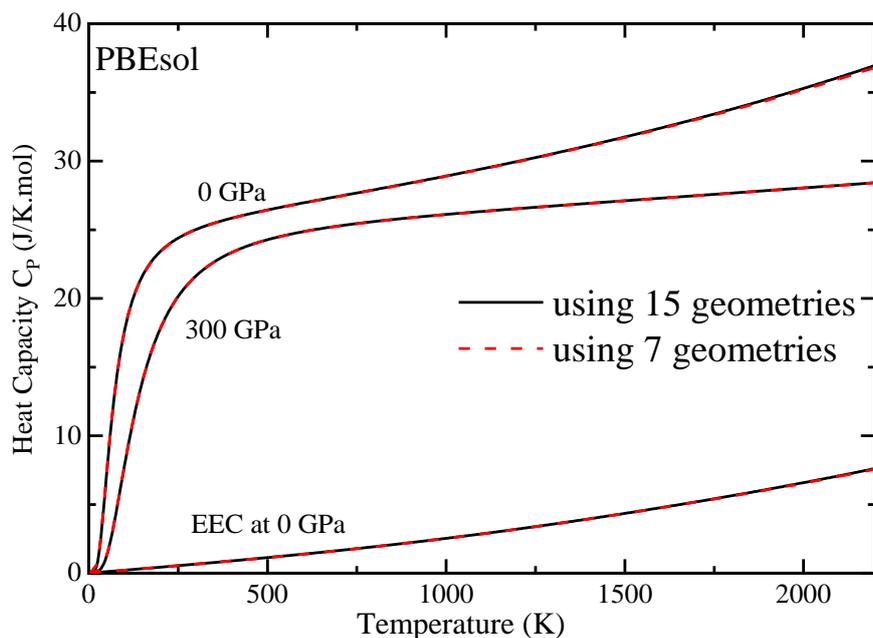

**Fig.S1C:** Temperature-dependent isobaric heat capacity ($C_P$) obtained for PBEsol at 0 GPa and 300 GPa evaluated on 15 and 7 geometries. The electronic excitation contribution (EEC) to the $C_P$ at 0 GPa is shown.

- Bulk modulus (B):

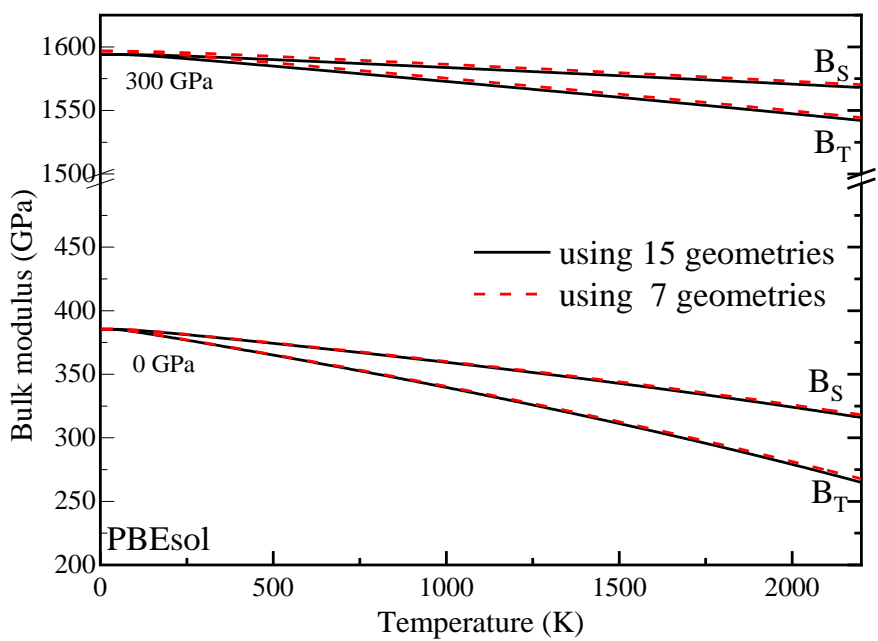

**Fig.S1D:** Temperature-dependent isoentropic ($B_S$) and isothermal ($B_T$) bulk modulus obtained for PBEsol at 0 GPa and 300 GPa evaluated on 15 and 7 geometries.

- Thermodynamic average Grüneisen parameter (γ):

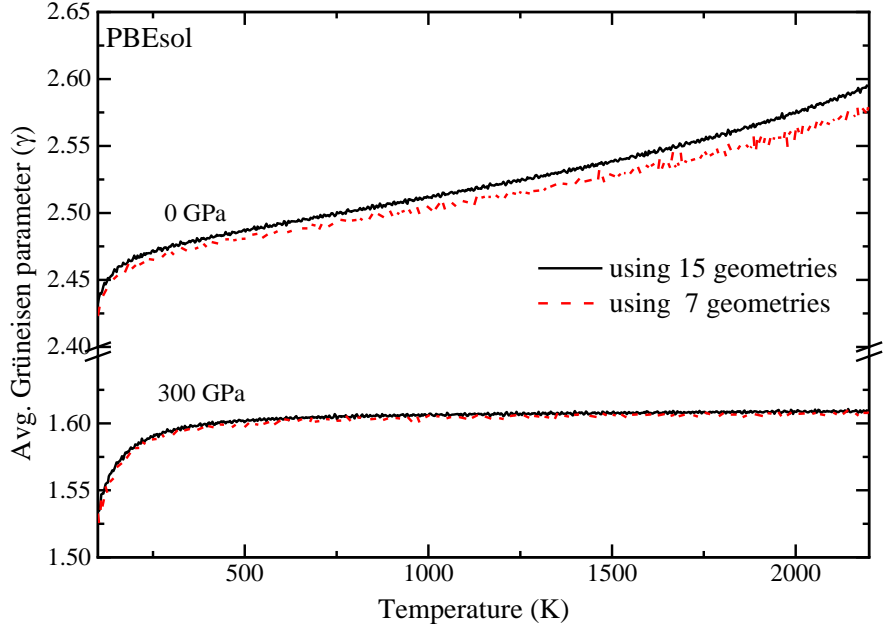

**Fig.S1E:** Temperature-dependent thermodynamic average-Grüneisen parameter obtained for PBEsol at 0 GPa and 300 GPa evaluated on 15 and 7 geometries.

**S2: Comparison between phonon dispersion obtained using different k-point and q-point mesh:**

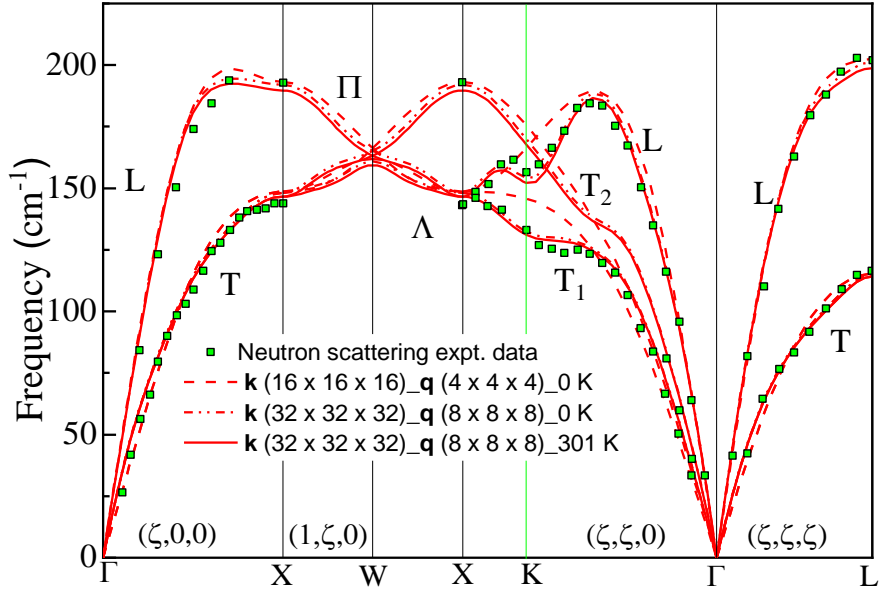

**Fig.S2:** Phonon dispersions curve obtained at 0 K (dashed) using **k**-points 16 × 16 × 16, **q**-points mesh 4 × 4 × 4, and **q**-points mesh, and at 0 K (dash-dot-dot) and 301 K (solid lines) using **k**-points 32 × 32 × 32, **q**-points mesh 8 × 8 × 8 and **q**-points mesh. For comparison, the experimental neutron in-elastic scattering data is included.

The phonon dispersion interpolated at T=0 K using 16 × 16 × 16 **k**-points mesh and 4 × 4 × 4 **q**-points mesh is presented in Fig.S3. For comparison, we included the phonon dispersion interpolated at T = 0 K and 301 K (shown in the main article) using 32 × 32 × 32 **k**-points mesh and 8 × 8 × 8 **q**-points mesh, and experimental inelastic scattering data. From Fig.S3, it is evident that the choice of 32 × 32 × 32 **k**-points mesh and 8 × 8 × 8 **q**-points mesh successfully explains the dispersions between **X** to **K** to **Γ**, which is not described well using 16 × 16 × 16 **k**-points mesh and 4 × 4 × 4 **q**-points mesh.